\documentclass[doublecol]{epl2} 
\usepackage{graphicx}
\def\lsim{\mathrel{\rlap{\lower4pt\hbox{\hskip1pt$\sim$}}
    \raise1pt\hbox{$<$}}}                
\def\gsim{\mathrel{\rlap{\lower4pt\hbox{\hskip1pt$\sim$}}
    \raise1pt\hbox{$>$}}}                
\title{Self-organising mechanism of neuronal avalanche criticality}
\shorttitle{SOC in neuronal networks} 

\author{D. E. Juanico\thanks{E-mail: \email{djuanico@gmail.com}}}
\shortauthor{D. E. Juanico}

\institute{                    
  Rm 3111 Complex Systems Theory Group, National Institute of Physics \\
University of the Philippines - 
Diliman, Quezon City 1101, Philippines
}

\abstract{
A self-organising model is proposed to explain the criticality
in cortical networks deduced from recent observations of neuronal
avalanches. Prevailing understanding of self-organised criticality (SOC)
dictates that conservation of energy is essential to its emergence.
Neuronal networks however are inherently non-conservative as demonstrated
by microelectrode recordings. The model presented here shows that SOC can
arise in non-conservative systems as well, if driven internally. Evidence
suggests that synaptic background activity provides the internal drive for non-conservative
cortical networks to achieve and maintain a critical state. SOC is robust to
any degree $\eta \in (0,1]$ of background activity when the network size $N$ 
is large enough such that $\eta N\sim 10^3$. For small networks, a strong background 
leads to epileptiform activity, consistent with neurophysiological knowledge
about epilepsy.}

\begin{document}

\maketitle

Neuronal networks have been demonstrated to exhibit a type of activity dubbed
as ``neuronal avalanche"~\cite{BeggsPlenz2003}. A neuronal avalanche is 
characterised by a cascade of bursts of local field potentials (LFP), which
originate from synchronised action potentials triggered by a single neuron.
The intensity of the recorded LFP is assumed to be proportional to the 
number of neurons synchronously firing action potentials within a short 
time interval, and thus is a good measure of neuronal avalanche size~\cite{BeggsPlenz2003}.
The LFP bursts are brief compared to the observation period, typically lasting
tens of milliseconds and separated by periods of quiescence that last
several seconds. When observed with a multielectrode array, the number of
electrodes detecting LFPs, which is in turn proportional to LFP intensity,
is distributed approximately as an inverse power law with exponent $\approx 1.5$.
The observed power law has been intuitively linked to the principles of self-organised
criticality~\cite{BeggsPlenz2003} through results from the study of critical
branching processes~\cite{ZapperiLauritsenStanley1995}, and has been proposed
and modelled~\cite{HaldemanBeggs2005,Beggs2006} to consequently enhance the
information processing capability of cortical networks in vivo. It has been
experimentally demonstrated however that propagation of neural activity does
not conserve information content, which is encoded in the frequency of spike 
firing~\cite{VogelsAbbott2005}. This experiment sustains previous findings that
synaptically connected pairs of neurons exhibit information transmission failures
wherein action potentials from the pre-synaptic neuron do not {\it depolarise} 
the post-synaptic neuron~\cite{GalarretaHestrin1998}. Information loss during
transmission seems to preclude the observed neural synchrony in neuronal networks
through avalanche activity. Moreover, the critical branching process model concludes
that any degree of loss or non-conservation frustrates criticality and introduces
a characteristic size to the avalanche size distribution~\cite{LauritsenZapperiStanley1996}.
Thus until now a gap exists between experiment and theory in terms of explaining how cortical
networks maintain criticality despite inherent non-conservation in neuronal transmission.

Neurons receive and transmit information using one form of medium---the electric potential,
which is both received as input (synaptic potential) and fired as output (action potential
or local field potential). Each neuron stores the input as membrane potential and when
this exceeds a threshold the neuron fires. Usage of one form of medium in receiving, 
storing and transmitting information, as well as the all-or-none response of neurons
parallels with that of SOC sandpile models. A ``sand grain" serves as the currency of
exchange and a sandpile site only transfers grains when the amount of stored grains 
exceeds a particular threshold. Thus, neuronal avalanches may parsimoniously, yet 
sufficiently, be analysed using SOC sandpile models.

The large number of neurons and the high density of non-local synaptic connections (i.e.,
linking neurons distally located from each other) that comprise the brain~\cite{Purves2004}
allows the approximation of the physics underlying its information transport by mean-field 
models. A convenient  mean-field sandpile model that appropriately captures neuronal avalanche dynamics is the
self-organised critical branching process (SOBP) model introduced by Zapperi, Lauritsen
and Stanley~\cite{ZapperiLauritsenStanley1995}, which represents the avalanche as a
branching process. SOC in this model is closely connected with the criticality of the 
branching process for which the theory is well-established~\cite{Harris1963}. In the 
branching process theoretical framework, activity at one site generates subsequent activity in a number of
other sites. When the number of subsequently activated sites---the branching parameter
$\sigma$---is equal to unity on the average, the branching process is critical. Criticality
in the SOBP model has been achieved only when grain transfer during an avalanche is
conservative. This has been proven by incorporating into the model a probability $\epsilon$
that grains which dislodge from a toppling activated site is absorbed (or dissipated) rather
than exchanged. For any $\epsilon>0$, the branching parameter $\sigma<1$, and a 
characteristic size in the avalanche size distribution, which scales with $\epsilon^{-2}$,
is introduced. Avalanches are not sustained when the conservation law is violated; they
easily die out after a few topplings. The characteristic size diverges when $\epsilon\rightarrow 0$,
or when the SOBP system is conservative~\cite{LauritsenZapperiStanley1996}. Thus, representation
of neuronal avalanches, which display criticality while occurring in a non-conservative
substrate, through the SOBP model seems to be a contradiction~\cite{Juanico2006}. 

A plausible source of sustaining the drive to neuronal avalanches is synaptic background
activity~\cite{HoDestexhe2000}, which has been found to enhance responsiveness of neocortical
pyramidal neurons to sub-threshold inputs. Synaptic background activity occurs in the form
of membrane potential fluctuations. Because of these voltage fluctuations, small excitatory
inputs are able to generate action potentials in neurons. The enhancement of weak signal
detection capability in the presence of background activity is analogous to a well-studied
nonlinear phenomenon in physics known as \emph{stochastic resonance}. It has been suggested
that the level of background activity within actual in vivo cortical networks is optimal
and possibly keeps neurons in a highly responsive state~\cite{HoDestexhe2000}. In this Letter,
a phenomenological model of synaptic background activity is incorporated into the SOBP model
to realise SOC in the presence of a violation of the transmission conservation law. The non-conservative 
SOC model proposed here parallels the theoretical work of Pruessner and Jensen on a 
sandpile-like model defined to approach the random-neighbour forest-fire model
in the non-conservative limit~\cite{PruessnerJensen2002}.

A crucial starting point for describing the model is translating the sandpile language into
terms that define neuronal phenomena, done in the following. 
A site corresponds to a neuron, and its height is analogous
to the membrane potential. At ``rest," the membrane potential is typically 
$-90\un{mV}\leq V_{\rm m}\leq -40\un{mV}$~\cite{Purves2004}. A neuron can undergo a critical 
state determined by a threshold potential which is less negative than the resting potential,
but varies across different types of neurons. When the membrane potential crosses this threshold,
an action potential is initiated. These three neuronal states are mapped to sandpile height $z$
in the following manner: $z=0$ (resting state); $z=1$ (critical state); and $z=2$ (excited state).
Neurons only fire action potentials when they are in the excited state. These mappings also
agree with the definition of the Manna sandpile~\cite{Manna1991}. Lastly, grain addition and
subtraction operations in sandpile models correspond to depolarisation and hyperpolarisation,
respectively. Depolarisation displaces the membrane potential of a neuron towards a less 
negative value, while hyperpolarisation displaces the membrane potential towards a more
negative value~\cite{Purves2004}. In the model, depolarisation corresponds to a change
$\Delta z=+1$ in the membrane potential whereas hyperpolarisation to a change $\Delta z=-1$ or $-2$.
Albeit the actual values of $V_{\rm m}$ are continuous, the discretisation employed here is 
sufficient to emulate the crucial role played by the threshold potential to neuronal activation.

The model cortical network is assumed to be a fully-connected random network in order to simplify
the mean-field treatment. Although fully-connected, only two randomly-chosen synaptic connections 
of a neuron are potentially utilised in every avalanche event. 
The network has a size of $N=2^{n+1}-1$ neurons, 
where $n$ is the  upper bound on the number of depolarisations (caused by action potentials triggered by a
single neuron) that can take place in a single avalanche. This serves as the boundary condition
to a mean-field model that neglects spatial details~\cite{ZapperiLauritsenStanley1995}. The 
network is in a ``quiescent" state when no excited neurons are present;
otherwise it is ``activated." The densities of critical neurons and resting neurons in a 
quiescent network are $\rho$ and $1-\rho$, respectively. The network is slowly stimulated
by external stimuli. The probability that the stimulus excites a critical neuron simply corresponds
to the density $\rho$. At this point, the network is activated and avalanche ensues. The excited
neuron fires an action potential that is transmitted to two post-synaptic neurons, chosen at
random. Three different things may happen: (i) with probability $\alpha$ the excited neuron hyperpolarises
to resting state ($z:2\rightarrow 0$) and both post-synaptic neurons depolarise; (ii) with probability
$\beta$ the excited neuron hyperpolarises to critical state ($z:2\rightarrow 1$) and only one of the
two post-synaptic neurons depolarises; and (iii) with probability $\epsilon=1-\alpha-\beta$ the
excited neuron hyperpolarises to resting state ($z:2\rightarrow 0$) but the action
potential fails to transmit thereby none of the post-synaptic neurons depolarise. These rules are
applied iteratively for every excited neuron that emerges until the whole network recovers its
quiescent state or when the action potential propagates a total of $n$ steps, after which no
subsequent depolarisation takes place. The $n$-th neuron, which serves as the boundary of 
avalanche propagation, may loosely be interpreted as a peripheral or a motor neuron.
In the wake of an avalanche, the density $\rho$ of the quiescent network changes.
Avalanches take place instantaneously with respect to the period that the network lingers in the 
quiescent state. This timescale separation is also observed experimentally~\cite{BeggsPlenz2003}, 
and is evident in the fast processing time
of the brain when presented with a sensory stimulus~\cite{Purves2004}. 

Treating the neuronal avalanche as a branching process allows us to define a branching probability
\begin{equation}\label{BranchingProbability}
\pi_k = \alpha\rho\delta_{k,2}+\beta\rho\delta_{k,1}+\left[1-(1-\epsilon)\rho\right]\delta_{k,0}\, ,
\end{equation}
where $k$ denotes the number of post-synaptic neurons that get depolarised by the action potential
released by an excited neuron. The first term of Eq.~(\ref{BranchingProbability}) represents case (i),
the second term represents case (ii), and the last term is the sum of the probability described in
case (iii) and the probability $(1-\rho)$ that the neuron subsequently depolarised is a resting
neuron, which also yields no action potential. Having $\epsilon>0$ (or equivalently $\alpha+\beta<1$) makes
the sandpile model violate the transmission conservation law. A branching parameter 
$\sigma=\left(2\alpha+\beta\right)\rho$ is calculated from Eq.~(\ref{BranchingProbability}) according
to the definition $\sigma=\sum_k k\pi_k$~\cite{Harris1963}. The branching process is sub-critical if
$\sigma<1$, whereby an avalanche typically dies out after a few ($<n$) transmission steps. The branching
process, on the other hand, is supra-critical if $\sigma>1$, describing an avalanche that percolates
the entire network with nonzero probability. At the critical value $\sigma=1$, the avalanche size
distribution $P(s)$ is calculated using a generating function directly derivable from Eq.~(\ref{BranchingProbability}):
\begin{equation}\label{GeneratingFunction}
F(\omega) = \frac{1-b\omega-\sqrt{1-2b\omega+a\omega^2}}{2\alpha\rho\omega}\, .
\end{equation}
where $a=\beta^2\rho^2-4\alpha\rho\left[1-(1-\epsilon)\rho\right]$ and $b=\beta\rho$, and is related
to $P(s)$ as a complex-variable power series
\begin{equation}\label{PowerSeries}
F(\omega) = \sum_{s=1}^N P(s)\omega^s\, .
\end{equation}
Taking $N\rightarrow\infty$, expansion of Eq.~(\ref{GeneratingFunction}) about the singularity $\omega=0$
followed by the comparison of terms to the coefficients of Eq.~(\ref{PowerSeries}) results to a
recurrence relation of $P(s)$ valid for $s\geq 2$:
\begin{equation}\label{SizeDistribution}
P(s) = \frac{bP(s-1)(2s-1)-aP(s-2)(s-2)}{s+1}\, ,
\end{equation}
and subject to the end conditions: $P(0)=0$ and $P(1)=(b^2-a)/4\alpha$. At the critical state, $a=2b-1$ or
equivalently, the function
\begin{equation}\label{CriticalityCondition}
Q(\rho) = \beta^2\rho^2-4\alpha\rho\left[1-(1-\epsilon)\rho\right]-2\beta\rho + 1
\end{equation}
is equal to zero, which follows from $\sigma=1$. Eq.~(\ref{SizeDistribution}) asymptotically approaches
power-law behaviour with exponent $3/2$ for $s>>1$, which can be shown graphically. A closed form
solution for~Eq.~(\ref{SizeDistribution}) can also be easily solved by setting $\beta=0$~\cite{PinhoPrado2003}.

The phenomenological model for synaptic background activity is conceived by assuming the presence of
sub-threshold membrane potential fluctuations in neurons when the network is in the quiescent state.
The background activity level is quantified by a parameter $\eta\in(0,1]$. A resting neuron depolarises
with probability $\eta$ to become critical ($z:0\rightarrow 1$); otherwise it remains in the resting
state. Parallel update of all neurons through this process contributes a 
change $\eta(1-\rho)$ to the density $\rho$ of critical neurons.
This process represents the stochastic fluctuations arising from excitatory post-synaptic potentials (EPSPs) on the
membrane potential~\cite{Purves2004}. On the other hand, a critical neuron hyperpolarises to resting
state ($z:1\rightarrow 0$) with probability $\eta$ multiplied by the deviation of the branching
parameter per critical neuron $\sigma/\rho$ from unity. Parallel update over all neurons through
this process contributes a change $-\eta(\sigma/\rho-1)\rho$ to the density $\rho$, and represents
the stochastic fluctuations arising from inhibitory post-synaptic potentials (IPSPs) on the membrane
potential~\cite{Purves2004} and the feedback mechanism to suppress runaway excitation~\cite{NelsonTurrigiano1998}. 
Both effects of EPSPs and IPSPs operate on the same timescale and can be conveniently
added together to give a net change $\Gamma = \eta(1-\sigma)$ to the density $\rho$ of critical
neurons in the quiescent network. $\Gamma$ is dominantly excitatory to the network when $\sigma<1$,
which effectively increases $\rho$; on the other hand, it is dominantly inhibitory when $\sigma>1$,
which effectively decreases $\rho$. Experimental evidence shows that this type of seesaw
between cortical excitation and inhibition is a form of synaptic plasticity that contributes in
stabilising cortical networks, keeping them on the border between inactivity and epileptiform
activity~\cite{GalarretaHestrin1998}. The parameter $\eta$ takes on a more physical meaning
because experiments reveal an internal mechanism for differential synaptic depression 
that dynamically \emph{adjusts} the balance between cortical excitation and inhibition to foster
cortical network stability~\cite{NelsonTurrigiano1998}.

Summing the changes in $\rho$ brought about by synaptic background activity $\Gamma$ and 
the neuronal avalanche lays out the following dynamical equation
\begin{equation}\label{DynamicalEquation}
\frac{\upd\rho}{\upd t} = \eta[1-\sigma(\rho)]+A(p)+\frac{\xi}{N}\, .
\end{equation}
The first term is $\Gamma$ and the second term represents the change in $\rho$ in the wake
of a neuronal avalanche,
\begin{equation}\label{Avalanche}
A(\rho) = \frac{1}{N}\left\{1-\sigma^n-\frac{\epsilon\rho}{1-(1-\epsilon)\rho}
\left[1+\frac{1-\sigma^{n+1}}{1-\sigma}-2\sigma^n\right]\right\}\,,
\end{equation}
which is derived by following closely the branching process arguments of the analysis
in~\cite{LauritsenZapperiStanley1996}. The last term in~(\ref{DynamicalEquation}) 
represents the fluctuations,
which properly vanish in the limit of large $N$, around average quantities assumed
in the calculation of $\Gamma$ and $A$. This noise term is therefore neglected in 
the mean-field approximation for large $N$. The first term inside $\left\{\cdot\right\}$
of~(\ref{Avalanche}) represents the depolarisation caused by external stimuli
that consequently brings a critical neuron to excited state and initiate an action
potential. The second term represents the average amount of action potential dissipated
by the $n$-th neuron. The third term is present when $\epsilon>0$ and represents the
average amount of action potential absorbed due to transmission failure. Hence,
$\epsilon$ is the degree of non-conservation in the cortical network.

\begin{figure}[ht]
\onefigure[width=80mm]{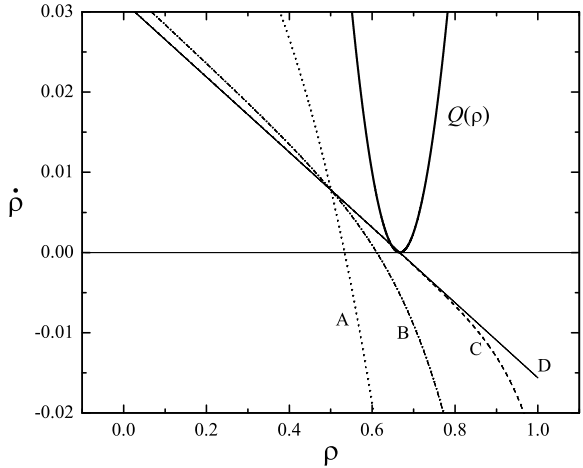}
\caption{Phase portrait of the density $\rho$ of critical neurons for a network with degree
of non-conservation $\epsilon=0.25$ driven by a background with $\eta=0.03125$ at different
sizes $N$: (A) $N=31$, dotted curve; (B) $N=511$, chain curve; (C) $N=131071$, dashed curve; and 
(D) $N\rightarrow\infty$, full curve. Also plotted is the function $Q(\rho)$ 
with root at $\rho_{\rm c}=\frac23.$}
\label{Figure1}
\end{figure}

The stationary behaviour of Eq.~(\ref{DynamicalEquation}) is analysed in phase space.
Fig.~\ref{Figure1} plots the phase portraits of $\rho$ for different network sizes
$N$. All networks have the same degree of non-conservation and background intensity, 
$\epsilon=0.25$ and $\eta=0.0625$, respectively. The fixed points $\rho^*$ are the 
roots of the phase portraits. Also shown is the function $Q(\rho)$ defined in
Eq.~(\ref{CriticalityCondition}) with a unique root at $\rho_{\rm c}=2/3$. As $N$
increases, $\rho^*\rightarrow\rho_{\rm c}$. However, even for a network as small
as $N=131071$ neurons, $|\rho^*-\rho_{\rm c}| << 1$. The rapid convergence
of $\rho^*$ to $\rho_{\rm c}$ from $N=31$ to $N=131071$ and the diminishingly 
perceptible difference between the phase portraits from $N=131071$ to $N\rightarrow\infty$
indicate a phase transition.

\begin{figure}[ht]
\onefigure[width=80mm]{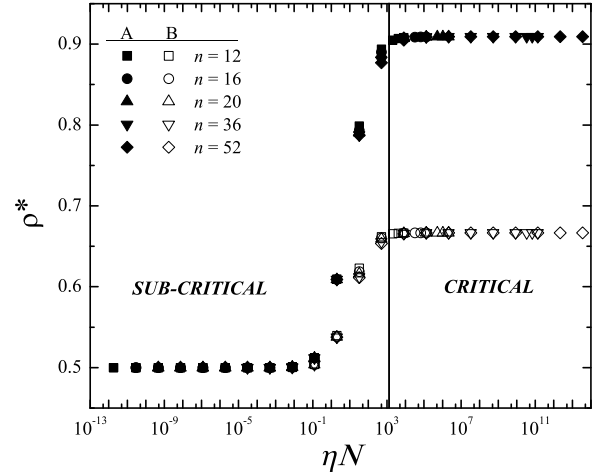}
\caption{Fixed point $\rho^*$ versus $\eta N$ exhibiting data collapse for networks of 
varying sizes with two different degrees of non-conservation: 
(A) $\epsilon=0.45$, $\rho_{\rm c}=0.909$, filled polygons; and
(B) $\epsilon=0.25$, $\rho_{\rm c}=0.667$, open polygons. Transition occurs between
sub-critical phase ($\rho^* <\rho_{\rm c}$) and critical phase ($\rho^*=\rho_{\rm c}$).
Phases are separated by a solid line footed at $\eta N\sim 10^3$.}
\label{Figure2}
\end{figure}

Indeed, as shown in Fig.~\ref{Figure2}, a transition from sub-critical phase ($\rho^*<\rho_{\rm c}$)
to critical phase ($\rho^*=\rho_{\rm c}$) occurs. Data collapse reveals that the profile of
the phase transition does not depend on $\eta$ and $N$ separately, but rather on the product $\eta N$.
In the sub-critical phase, the quiescent network has $\rho=\frac12$ of critical neurons,
in accordance with the results of~\cite{LauritsenZapperiStanley1996}. But starting at 
$\eta N\sim 10^{-1}$, the network driven by a background activity abruptly transforms to a 
critical phase up to $\eta N\lsim 10^{3}$. The critical phase is characterised by
$\rho=\rho_{\rm c}=(2\alpha+\beta)^{-1}$ (or $\rho_{\rm c}=[2(1-\epsilon)]^{-1}$ for $\beta=0$) 
for which the branching parameter $\sigma=1$.
Starting at $\eta N\sim 10^3$, the network is in the critical phase for any degree of
non-conservation $\epsilon\in(0,0.5]$ (at $\epsilon>0.5$, the critical density $\rho_{\rm c}>1$, hence
impossible to achieve). The phase transition implies that if the network is very large ($\sim 10^{12}$ neurons), 
which is typical of actual cortical networks~\cite{Purves2004}, 
then a wide range of $\eta$ enables the network to maintain its
critical phase, as long as $\eta N\geq 10^3$. However, since the fluctuations $\xi/N$ are neglected in 
analysing the stationary behaviour of Eq.~(\ref{DynamicalEquation}), this conclusion does not necessarily
hold true for networks with intermediate sizes, for which the fluctuations may not be negligible. 
Thus, simulation results are essential in probing the stationary behaviour of the network for this case.

\begin{figure}[ht]
\centering
\includegraphics[width=80mm]{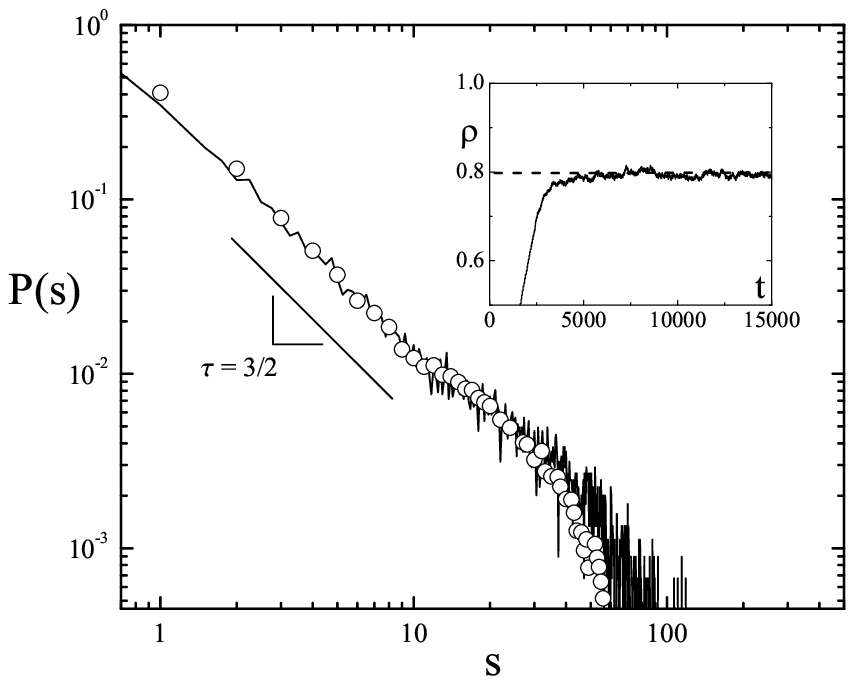}
\includegraphics[width=80mm]{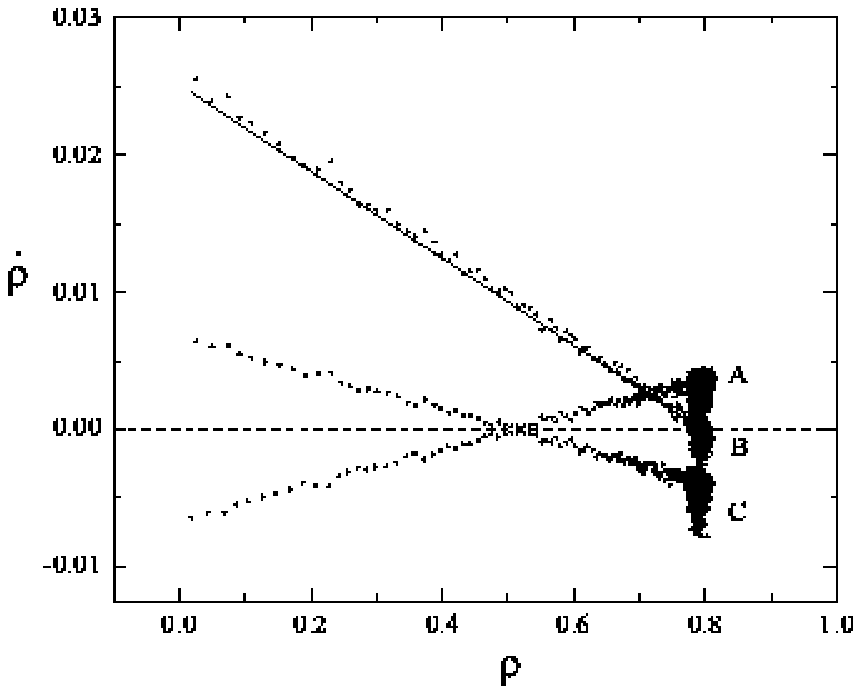}
\caption{{\it Top panel}.---Distribution of neuronal avalanche size data adopted from~\cite{HaldemanBeggs2005}
(circles), and of a simulated network (full-curve) with $\eta N\approx 3670$ having a degree
of non-conservation $\epsilon=0.25$. Inset graph shows $\rho$ versus iterations $t$ approaching
a critical value $\rho_{\rm c}=0.8$ in the steady state. {\it Bottom panel}.---Phase plot of the 
simulated network, starting at $\rho=0$, converging towards three dynamical attractors: (A) excitatory
background fluctuations; (B) fixed point ($\rho^*=\rho_{\rm c}$); 
and (C) inhibitory background fluctuations. The solid line is
the mean-field prediction.}
\label{Figure3}
\end{figure}

A simulation of the model is performed for a network with size $N=131071$ driven by
a synaptic background with $\eta=0.025$ such that $\eta N\approx 3670$. 
The degree of non-conservation in the simulated network is $\epsilon=0.25$ such that
the critical density $\rho_{\rm c}=0.8$, in accord with the findings of Vogels and
Abbott that depolarisations due to synchronous action potentials 
evoke post-synaptic spikes only $80\%$ of the times~\cite{VogelsAbbott2005}. After
rescaling the simulated avalanche size with a factor deduced from data in~\cite{BeggsPlenz2003},
the simulated network successfully fits the experimental data for the neuronal avalanche
distribution adopted from~\cite{HaldemanBeggs2005}, as shown in Fig.~\ref{Figure3} ({\it Top panel}). 
A power-law with exponent $3/2$ mainly characterises the distribution. However, an 
exponential cutoff appears because
of the limited number of microelectrodes used to resolve LFP intensity during the 
experiments~\cite{BeggsPlenz2003}. The cutoff shifts to the right (i.e., towards larger avalanche
sizes) when the number of microelectrodes is increased. The simulation data also fits this
exponential tail, which also arises from the boundary condition that effectively puts an 
upper bound $n$ to the number of transmission steps during an avalanche. The cutoff also
shifts to the right by increasing $n$. Hence, both model and experimental data agree not 
only in the power-law behaviour of the avalanche size distribution, but also in the 
mechanism that gives rise to the exponential tail. The inset graph illustrates the evolution of 
the density $\rho$ in the quiescent network with time $t$, approaching the critical value 
$\rho_{\rm c}=0.8$ in the steady state. This steady-state behaviour is supported by a phase
plot of $\rho$ (Fig.~\ref{Figure3}, {\it Bottom panel}). The fixed point (B) corresponds to
$\rho_{\rm c}=0.8$. Also shown are two other dynamical attractors (A, C), which correspond
to background activity fluctuations neglected in the fixed-point
analysis of Eq.~\ref{DynamicalEquation}. The symmetry between (A) and (C) evidently suggests
the balance of cortical excitation and inhibition, imparting stability to the network at
the critical state. Thus, the background activity is essential in maintaining the network
at the critical state such that a power-law avalanche size distribution is generated.

\begin{figure}[ht]
\onefigure[width=80mm]{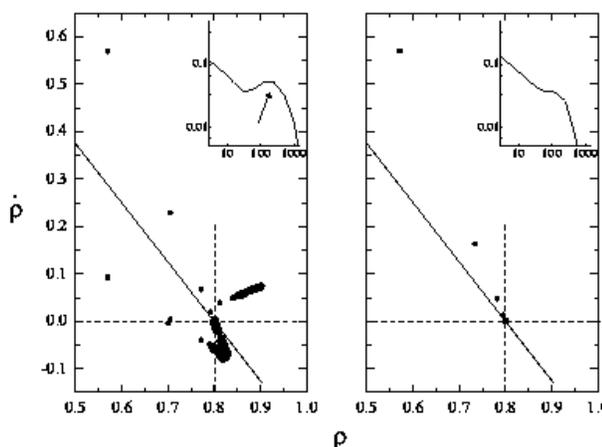}
\caption{Phase plots for two networks of different sizes $N$ but
similar degrees of non-conservation $\epsilon=0.25$ and
driven by a strong background $\eta=1.0$. Critical density is at $\rho_{\rm c}=0.8$. 
{\it Left panel}.---$N=131071$ neurons,
showing a prominent protrusion at $\rho>\rho_{\rm c}$ and $\dot{\rho}>0$. Solid
line is the mean-field prediction. Inset graph
displays the logarithmically-binned distribution of avalanche sizes with a marked peak
(pointed by arrow) for large sizes, indicating epileptiform activity. 
{\it Right panel}.---$N=4194303$ neurons, showing
stability of the fixed point at $\rho_{\rm c}$ consistent with mean-field
prediction (solid line). Inset graph displays the logarithmically-binned distribution
of avalanche sizes exhibiting considerable power-law behaviour.}
\label{Figure4}
\end{figure}

Actual neuronal networks are found to be redundant---several identical neurons that perform similar
roles are present~\cite{Purves2004}. Redundancy thereby enlarges the network, and may
be vital to its robustness. Through the model, robustness of the critical behaviour due to network 
size is explored by driving a small ($N=131071$) and a large ($N=4194303$) network 
with a strong background ($\eta=1.0$). Fig.~\ref{Figure4} ({\it Left panel}) shows a small network
driven by strong background exhibiting an asymmetry in the excitatory and inhibitory dynamical
attractors---there is a larger region of excitation than inhibition. This mechanism is believed to
be the precursor of epileptic seizures, which manifests in the synchronous firing of a large
number of neurons. The inset graph illustrates the avalanche size distribution for this network.
A marked peak appears (indicated by arrow) for large avalanche sizes of the distribution. Hence, there is
a high probability for a large number of neurons to synchronously depolarise and fire action 
potentials. This feature indicates that small networks show epileptiform activity when driven by a 
strong background. This is consistent with neurophysiological knowledge that seizure development is
triggered by damages to brain cells caused by injury, drug abuse, degenerative neurodiseases,
brain tumors, and brain infections~\cite{Purves2004}. On the other hand, the large network is 
robust to strong background activity (Fig.~\ref{Figure4},  {\it Right panel}). The avalanche size 
distribution of this network (inset graph) exhibits the expected power-law behaviour and there is 
no pronounced peak for large avalanche sizes. 

{\it Summary and Conclusion}.---A self-organising mechanism for neuronal avalanche activity is
proposed. Criticality manifests in the power-law behaviour of neuronal avalanche sizes. Despite
an inherent transmission non-conservation in neuronal networks, this criticality is maintained.
A large neuronal network that is internally driven by synaptic background activity self-organises 
towards and robustly maintains a critical state for any level of background activity $\eta\in (0,1]$ 
and for any degree of non-conservation $\epsilon\in(0,0.5]$. This finding advocates the role of
redundancy of neuronal networks in fostering robustness against any fluctuations of internal
activity. A small neuronal network, on the other hand, loses the balance between cortical 
excitation and inhibition when driven by a strong background ($\eta=1.0$), consequently leading 
to epileptiform activity. This finding agrees with neurophysiological knowledge that brain cell
damage is a chief contributor to the onset of seizure development.

The model also proves that self-organised criticality can be achieved even when a conservation law
of dynamical transfer is violated. A transition occurs from a sub-critical to a critical phase,
demonstrating a vast regime for non-conservative systems to display SOC behaviour.
Thus, it addresses a long-standing issue that is fundamental to the theory of self-organised 
criticality---whether conservation is necessary for its emergence.

{\it Recommendations}.---A key assumption in the model is network randomness.
However, the morphology of actual neuronal networks may actually be more 
accurately characterised in terms of small-world connectivity patterns~\cite{Goh2003,Roxin2004,Lee2006}. 
Thus, an investigation into the behaviour of the model in small-world networks is
recommended as a possible extension of this study. Nevertheless, the main conclusions,
which do not strongly depend on the network morphology as long as the density of 
connections and the size of the network remain large, would still hold.





\acknowledgments
The author wishes to acknowledge the Office of the Vice-Chancellor for Research and 
Development (OVCRD) of the University of the Philippines, Diliman for Research Grant
No. 050501 DNSE and the Philippine Council for Advanced Science and
Technology Research and Development (PCASTRD) for funding.

\end{document}